\documentclass[aps,pra,floatfix,twocolumn,superscriptaddress,nofootinbib, showpacs]{revtex4} 

\usepackage{graphicx}
\usepackage{graphics}
\usepackage{amssymb}
\usepackage{amsmath}
\usepackage{epsfig}
\usepackage{latexsym}
\usepackage{color}

\newcommand{\bra}[1]{\langle{#1}|}
\newcommand{\ket}[1]{|{#1}\rangle}

\def\ketc[#1]{\vert #1 \rangle}
\def\brac[#1]{\langle #1 \vert}

\newcommand{\beq}{\begin{equation}}
\newcommand{\eeq}{\end{equation}}
\newcommand{\bqa}{\begin{eqnarray}}
\newcommand{\eqa}{\end{eqnarray}}
\newcommand{\nn}{\nonumber}

\newcommand{\dg}{^\dagger}

\begin{document}

\title{A Practical Scheme for Error Control using Feedback}

\author{Mohan Sarovar}
\email{mohan@physics.uq.edu.au} \affiliation{Centre for Quantum
Computer Technology, and School of Physical Sciences, The
University of Queensland, St Lucia, QLD 4072, Australia}

\author{Charlene Ahn}
\email{cahn@theory.caltech.edu}
\affiliation{Institute for Quantum
Information, California Institute of Technology, Pasadena, CA
91125, USA}

\author{Kurt Jacobs}
\email{k.jacobs@griffith.edu.au} \affiliation{Centre for Quantum
Computer Technology, Centre for Quantum Dynamics, School of
Science, Griffith University, Nathan, QLD 4111, Australia}

\author{Gerard J. Milburn}
\email{milburn@physics.uq.edu.au} \affiliation{Centre for Quantum
Computer Technology, and School of Physical Sciences, The
University of Queensland, St Lucia, QLD 4072, Australia}


\begin{abstract}
We describe a scheme for quantum error correction that employs 
feedback and weak measurement rather than the standard tools of 
projective measurement and fast controlled unitary gates. The advantage 
of this scheme over previous protocols (for example Ahn et. al, PRA, 65, 042301 
(2001)), is that it requires little side processing while remaining robust to 
measurement inefficiency, and is therefore considerably more practical. ÊWe 
evaluate the performance of our scheme by simulating the correction of bit-flips. 
We also consider implementation in a solid-state quantum computation architecture and 
estimate the maximal error rate which could be corrected with current 
technology. 
\end{abstract}

\maketitle


\section{Introduction}
In the mere space of a decade, quantum information theory has blossomed
into a burgeoning field of
of experimental and applied research. The initial push
for this rapid development was provided by Peter Shor's discovery
of an algorithm that enables quantum computers to find the period
of a periodic function much more efficiently than any (known)
classical computer algorithm \cite{shor}. However, even after
Shor's discovery there was much doubt about the practicality of
quantum computing devices due to their fragile nature. The
coherencies between systems carrying the quantum information
that are crucial to quantum computing algorithms are extremely
vulnerable and easily destroyed by unavoidable interactions with
the surrounding environment. Furthermore, aside from this
\textit{decoherence}, another concern was the accumulation of
errors introduced by imperfect
operations performed on the encoded information.

Both these concerns were largely put to rest by the key development of
quantum fault tolerance. The error accumulation was shown to be
tolerable as long as the systematic error introduced by each
operational element was below a critical threshold value\cite{klz98}.
This threshold result relies heavily upon the development of quantum
error correction codes. These codes, the first of which were discovered
by Shor \cite{shor95} and Steane \cite{steane96}, redundantly encode
information in a manner that allows one to correct errors while
preserving coherencies and thus the encoded information.

The main ingredients in the implementation of these error control
codes are projective von Neumann measurements that discretize the
errors into a finite set, and fast controlled unitary gates that
provide the ability to correct any corrupted data. Of course,
instantaneous projective measurements and unitary gates are not
perfectly implementable in any system; here we will be concerned with
the details of how one would actually implement error correction
practically on a system where the physical tools available are
necessarily physically limited.

We extend previous work \cite{ahn-dl} and describe an
implementation of error control that utilizes stabilizer
error-correcting codes and employs weak measurement and Hamiltonian
feedback to effectively protect an unknown quantum state. This scheme
has similarities to the one described previously in \cite{ahn-dl}; however,
whereas that protocol uses a full state estimation technique that is
computationally intensive, this one uses a simple filtering technique
that is easily implementable. This protocol is therefore a reasonable
one for many of today's quantum computing architectures.

This paper is organized as follows. Section \ref{sec:rev} reviews the
key ideas we use: weak measurement, feedback, and stabilizer
codes. Section \ref{sec:scheme} describes our error control scheme and
two specific instances of it. Section \ref{sec:sims} describes the
simulations performed to analyze the performance of our scheme,
presents the results, and considers the effect of measurement
inefficiency. Section \ref{sec:real} examines an actual quantum
computer architecture and how this error control protocol could be
implemented on this architecture. Section \ref{sec:conc}
concludes.

\section{Review of concepts}\label{sec:rev}
\subsection{Quantum feedback control}
In order to describe the behavior of a quantum system with feedback,
we must first examine the description of an open quantum system
undergoing continuous weak measurement; later we will add feedback
conditioned on the measurement results. Continuous measurement is
modelled by considering the system of interest $\mathcal{S}$ to be
weakly coupled to a reservoir $\mathcal{R}$. In order to utilize the
Born-Markov approximation, we assume that the self-correlation time of
the reservoir is small compared to the time-scales of the
system-reservoir coupling and the system dynamics. This essentially
says that $\mathcal{R}$ measures $\mathcal{S}$ continuously but
quickly forgets, or dissipates away, the result of the
measurement. This allows us to write the unconditional dynamics of
$\mathcal{S}$ as the following master equation \cite{wiseman-thesis}:
\begin{equation}
\label{eq:master} \dot{\rho}(t)= -i[H,\rho(t)] + \sum_{\mu=1}^m
\kappa_\mu\mathcal{D}[c_\mu]\rho(t).
\end{equation}
Here $\rho$ is the reduced density matrix of $\mathcal{S}$, $H$ is
the system Hamiltonian, $\{c_\mu\}$ are the collection of
system-reservoir interactions (in our case, where we are
considering these interactions to be weak measurements, these are
the Hermitian operators corresponding to the observables),
$\kappa_\mu$ is a parameterization of the coupling strength of
$c_\mu$, and $\mathcal{D}$ is the decoherence superoperator given by
\begin{equation}
\label{eq:D} \mathcal{D}[A]\rho = A\rho A\dg - \frac{1}{2}(A\dg
A\rho + \rho A\dg A)
\end{equation}
for any operator $A$. Note that we set $\hbar=1$ throughout this
paper, except for section \ref{sec:real}.

This equation describes the unconditional dynamics of $\mathcal{S}$ because
we are assuming that the measurement records are ignored. That is,
ignoring the measurement records corresponding to the observables
$\{c_\mu\}$ means that the best description of $\mathcal{S}$ we
have is one where we average over all possible measurement
records, and hence all possible \textit{quantum trajectories} that
the system could have traversed \cite{wiseman-thesis, wiseman-fb}.

If we choose not to ignore the measurement records,
we instead get a conditional evolution equation for the system
\cite{wiseman-thesis, wiseman-fb}:
\begin{eqnarray}
\label{eq:mastercond} d\rho_c(t)= -i[H,\rho_c(t)]dt +
\sum_{\mu=1}^m \kappa_\mu\mathcal{D}[c_\mu]\rho_c(t)dt \nn \\ +
\sum_{\mu=1}^m
\sqrt{\kappa_\mu}\mathcal{H}[c_\mu]\rho_c(t)dW_\mu(t)
\end{eqnarray}
where $\rho_c$ is the system density operator conditioned on the
measurement records of $\{c_\mu\}$, $dW_\mu(t)$ are \textit{Weiner
increments} (Gaussian distributed random variables with mean zero
and autocorrelation $\langle dW(s)dW(t)\rangle = \delta(s-t)dt$)
\cite{gardiner}, and $\mathcal{H}$ is the superoperator
\begin{equation}
\label{eq:H} \mathcal{H}[A]\rho = A\rho + \rho A\dg - \rho
~\text{tr}[A\rho + \rho A\dg]
\end{equation}
for any operator $A$. The measurement record from the measurement 
of $c_\mu$ is:
\begin{equation}
\label{eq:dQ} dQ_\mu (t) = \kappa_\mu\langle c_\mu +
c_\mu\dg\rangle_c dt + \sqrt{\kappa_\mu}dW_\mu(t)
\end{equation}
where $\langle a \rangle_c = \text{tr}(\rho_c a)$. In terms of
quantum trajectories, this equation corresponds to a diffusive unravelling
of the master equation given in Eq. (\ref{eq:master}). From here onwards, for
simplicity we will specialize to the case of $m=1$ in equations
(\ref{eq:master}) and (\ref{eq:mastercond}) (i.e. only one
$c_\mu$).

Given this model, we can now consider adding feedback to the
system. In general, the feedback will be a function of the entire
measurement record history. And if we use Hamiltonian feedback,
the conditional stochastic master equation (SME) \textit{with
feedback} becomes
\begin{eqnarray}
\label{eq:mastercond-fb} d\rho_c(t) &=& -i[H,\rho_c(t)]dt \nn \\
&& + \kappa\mathcal{D}[c]\rho_c(t)dt +
\sqrt{\kappa}\mathcal{H}[c]\rho_c(t)dW(t) \nn
\\ && -iR_Q(t)[F,\rho_c(t)]dt
\end{eqnarray}
where $R_Q(t)$ is some arbitrary function of the entire
measurement history $Q(t)$, and $F$ is the feedback Hamiltonian.
Note that all we have done is to add a Hamiltonian evolution term
whose strength is conditioned by a function of the measurement
record.

As shown in \cite{wiseman-thesis}, in the restricted case that
$R_Q(t)$ is a linear function of the measurement value at time $t$
only (i.e. $R_Q(t_0) = f_{linear}[Q(t_0)]$), we can simplify equation
(\ref{eq:mastercond-fb}) further, and derive a master equation for the
unconditional dynamics of the system. This restricted case is often
referred to as \textit{Markovian} feedback, and is considered in
connection with quantum error control in \cite{ahn-wm}.  However, in
general, when the feedback is conditioned by a current that is some
arbitrary function of the measurement history, it is not possible to
treat the evolution analytically, and numerical simulation is the only
recourse for solving (\ref{eq:mastercond-fb}). An important special case of this general feedback is one in which the function $R_Q(t)$ is designed to compute an estimate of the state $\rho_c(t)$ and output an appropriate feedback strength \cite{DJ99,DHJMT00}. We will refer to this as \textit{Bayesian} feedback, following reference \cite{wiseman-mw}. In \cite{ahn-dl} an almost full
state estimation is done en route to error control, and here we will consider a simpler and more
practical version of that scheme that performs only a partial state
estimation.

\subsection{Stabilizer codes}
\label{sec:stabcodes} In making continuous weak measurements
on our system, we would like to choose the measurements in such a
manner that we gather as much information about the errors as possible
while disturbing the logical qubits as little as possible. These
are exactly the conditions satisfiable by encoding the information
using a quantum error correcting code; and the powerful stabilizer
formalism \cite{gottesman, mikeandike} provides a way to easily
characterize many of these codes. We will restrict our attention to
these stabilizer codes and in this section provide a brief description
of the main result of the formalism and give an example. For more
detailed discussions, the reader is referred to \cite{gottesman} and
\cite{mikeandike}.

We begin by introducing the Pauli group
\begin{equation}
\label{eq:pauligroup} P_n = \{\pm 1, \pm i\} \otimes \{I, X, Y,
Z\}^{\otimes n}
\end{equation}
where $X, Y$, and $Z$ denote the Pauli operators $\sigma_x,
\sigma_y$, and $\sigma_z$ respectively. To simplify notation, we
will omit the tensor product symbol when notating members of $P_n$
(e.g. $ZZI \equiv Z\otimes Z\otimes I$).

Now, if we encode $k$ logical qubits in $n$ physical qubits ($n\ge
k$), then we can think of errors on our physical system as the action
of some subset $\{E_j\} \subset P_n$ \cite{mikeandike}. Thus, we would
like to choose our encoding in a manner that allows us to detect and
correct the action of those group elements in $\{E_j\}$. The main
result from the theory of stabilizer codes tells us about the
possibility of choosing such an encoding. It says that provided the
elements of $\{E_j\}$ satisfy a certain condition, it is always
possible to choose a codespace, C, that can be used to detect and
correct these error elements \cite{gottesman}. Furthermore, this
codespace has some special properties:
\begin{enumerate}
\item There exist a set of operators in $P_n$, called the \textit{stabilizer generators} and denoted by $g_1, g_2, ..., g_r$, such that every state in C is an eigenstate with eigenvalue +1 of all the stabilizer generators. That is, $g_i\ket{\psi} = \ket{\psi}$ for all $i$ and for all states $\ket{\psi}$ in C. Moreover, these stabilizer generators are all mutually commuting. 
\item The stabilizer code error correction procedure involves
simultaneously measuring all the stabilizer generators and then
inferring what correction to apply from the measurement results. The formalism states that the stabilizer measurement results indicate a unique correction operation.
\end{enumerate}

This result tells us that once we identify a set of one qubit errors in $P_1$ that we are concerned about, it is possible to choose a stabilizer codespace that can be used to protect the encoded information against such errors. The error detection-correction procedure involves measuring each of the stabilizer generators and then applying a correction corresponding to the results obtained from the stabilizer measurements. 

\subsection{Example: Three qubit bit-flip code}
\label{sec:ex_bitflipcode} A common error encountered in quantum
computing implementations is the bit flip. This type of error has
the effect of reversing the encoded qubit's value at random times.
That is, $\ket{\psi} \rightarrow X\ket{\psi}$ with probability
$p$, and $\ket{\psi} \rightarrow \ket{\psi}$ with probability
$1-p$ (where again, $X \equiv \sigma_x$, and $\ket{\psi}$ is a one
qubit state).

One encoding that protects against this type of error is
\begin{eqnarray}
\label{eq:tq_encoding}
\ket{0}_L \equiv \ket{000}_P \nn \\
\ket{1}_L \equiv \ket{111}_P
\end{eqnarray}
where the right hand side shows the physical encoding in three qubits
of the logical qubit value on the left hand side. That is, C $=
\textrm{span}\{\ket{000}, \ket{111}\}$. The stabilizer generators for
this codespace are the operators $ZZI$ and $IZZ$. The code can be used
to detect and correct any of the errors $XII, IXI$, and $IIX$. The
correction procedure involves measuring the two stabilizer generators
and then applying the appropriate correcting unitary according to the
rules of table \ref{tab:tq_table}, which corrects for the bit-flip
errors.

\begin{table}
\begin{tabular}{|c|c|c|c|}
  \hline
  \textbf{ZZI} & \textbf{IZZ} & \textbf{Error} & \textbf{Correcting unitary} \\
  \hline\hline
  +1 & +1 & None & None \\
  -1 & +1 & on qubit 1 & XII \\
  +1 & -1 & on qubit 3 & IIX \\
  -1 & -1 & on qubit 2 & IXI \\
  \hline
\end{tabular}
\caption{The three qubit bit-flip code. Note that each error results in a different sequence of stabilizer generator measurement results.} \label{tab:tq_table}
\end{table}

\section{The error correction scheme}
\label{sec:scheme}

\subsection{The general scheme}
\label{sec:scheme_general} Once information is encoded using a
quantum error-correcting code, conventional error control proposals use
projective measurements to measure the stabilizer generators and
fast unitary gates to apply the corrections if necessary. In such
schemes the detection-correction operation, which is initiated by
the projective measurement, happens at discrete time intervals,
and these intervals are chosen so that the average number of
errors within an interval is correctable. We will refer to such
implementations as \textit{discrete error correction} schemes
because of the discrete nature of the detection-correction
operation.

In this section we present a protocol that combines weak
measurements of the stabilizer generators with feedback to perform
\textit{continuous} error correction. Because of the encoding,
these measurements will be unobtrusive when the system is in the
codespace and will give error specific information when it is
not. However, the requirement of \textit{weak} measurements makes
the measurement currents described by (\ref{eq:dQ}) noisy, and
therefore ineffective for feedback conditioning. In order to use
the information from these measurements to condition the feedback,
we must smooth out some of the noise in the currents. The smoothing can be
easily done using a low-pass filter; however, such a filtering
process introduces its own complications. Specifically, such filtering makes it impossible to derive a
master equation describing the evolution because the
noise in the feedback signal at time $t$ is not independent of the
system state at time $t$. In essence, our smoothing procedure
makes Markovian feedback impossible, and leaves the alternative of
Bayesian feedback.

Now, full state estimation is a computationally expensive
procedure -- it most often involves solving an SME in real-time. In fact, as shown in \cite{ahn-dl}, the
resources needed to apply a full state estimation feedback procedure to quantum error control scale
\textit{exponentially} with the number of qubits in the stabilizer
code. Fortunately, we do not need to do a full state
estimation. Instead, the coarse-grained state estimate that the
stabilizer measurements provide --- whether the state
is in the codespace or not, and if not, how to correct back into
the codespace --- is precisely the information needed for error
control. That is, instead of estimating $\rho_c(t)$, we simply need to reliably identify the stabilizer generator measurement results (in the presence of noise) in order to place $\rho_c(t)$ inside or outside the codespace.
Furthermore, as seen in the example of section
\ref{sec:ex_bitflipcode}, this information is contained in the
signatures of the stabilizer generator measurements (whether they are plus or minus one), a quantity
that is fairly robust under the influence of noise. These
observations suggest that weak measurement and feedback can be
used to continuously detect and correct errors.

The general form of the error correcting scheme we propose is
similar to discrete error control, but with a few modifications to
deal with the incomplete information gained from the weak
measurements. The scheme can be stated in four steps:
\begin{enumerate}
\item Encode information in a stabilizer code suited to the errors
of concern. \item Continuously perform weak measurements of the
stabilizer generators, and smooth the measurement currents. \item
Depending on the signatures of the smoothed measurement currents, form
conditioning signals for feedback operators on each physical
qubit. These conditioning currents will be highly
non-linear functions of the measurement currents because the
conditional switching based on signatures is a non-linear
operation. \item Apply feedback Hamiltonians to each physical
qubit, where the strength of the Hamiltonians is given by the
conditioning signals formed in the previous step.
\end{enumerate}

Given $m$ stabilizer generators and $d$ errors possible on our system, the SME describing the evolution of a system
under this error control scheme is
\begin{eqnarray}
\label{eq:master_ecc} d\rho_c(t) &=& \sum_{k=1}^d
\gamma_k \mathcal{D}[E_k]\rho_c(t)dt \nn \\
&+& \sum_{l=1}^m \kappa\mathcal{D}[M_l]\rho_c(t)dt + \sqrt{\kappa}\mathcal{H}[M_l]\rho_c(t)dW_l(t) \nn \\
&+& \sum_{k=1}^d -iG_k(t)[F_k, \rho_c(t)]dt
\end{eqnarray}
where $\gamma_k$ is the error rate for error $E_k$, $\kappa$ is
the measurement strength (assumed for simplicity to be the same for all
measurements $M_l$), $F_k$ is the feedback Hamiltonian correcting
for error $E_k$, and $G_k$ is the feedback conditioning signal for
$F_k$. Each $G_k$ is a conditional function of the signatures of
all the smoothed stabilizer measurements, $\{M_l\}$.

Equation (\ref{eq:master_ecc}) has three parts to it: the first
line describes the effects of the error operators, the second line
describes the effects of the weak stabilizer generator
measurements, and the third line describes the effect of the
feedback. [Also note that we have set the system Hamiltonian, $H$
in (\ref{eq:mastercond-fb}), to zero.]

This general scheme is illustrated by the following examples. The
systems described by these examples are also the ones simulated in
section \ref{sec:sims}.

\subsection{Example: A toy model}
\label{sec:scheme_toy} This first example is somewhat artificial,
but serves as a good illustration of our protocol. The `codespace' we
want to protect is simply the state $\ket{0}$, the errors are random
applications of $X$, and the protocol gathers information by measuring
the stabilizer generator $Z$. Obviously this `code' cannot be used for
any information processing, but it is useful for investigating the
behaviour of our feedback scheme.

The dynamics of this system before the application of feedback are
described by the following SME:
\begin{eqnarray} d\rho_c(t) &=&
\gamma\mathcal{D}[X]\rho_c(t)dt +
\kappa\mathcal{D}[Z]\rho_c(t)dt \nn \\
&&+ \sqrt{\kappa}\mathcal{H}[Z]\rho_c(t)dW(t),
\end{eqnarray}
where $\gamma$ is the error rate and $\kappa$ is the measurement
rate. The measurement current has the form
\begin{equation}
\label{eq:oq_meas} dQ(t) = 2\kappa\langle Z \rangle_c (t)dt +
\sqrt{\kappa}dW(t),
\end{equation}

Now, the measurement of $Z$ reveals whether the systems is in the
`codespace' or not because
\begin{eqnarray}
Z\ket{0} = +1\ket{0} \nn \\
Z\ket{1} = -1\ket{1}.
\end{eqnarray}
However, we do not have direct access to $\langle Z\rangle_c$, but rather
only to the noisy measurement current (\ref{eq:oq_meas}). Therefore we must
smooth out the noise on it to obtain error information, and we will choose the following simple filter to do so:
\begin{equation}
\label{eq:oq_filt} R(t) = \frac{1}{\mathcal{N}}\int_{t-T}^t
e^{-r(t-t')}dQ(t')
\end{equation}
This integral is a convolution in time between the measurement
signal and an exponentially decaying signal. In frequency space,
this acts as a low pass filter, and thus the output of this
operation is a smoothed version of the measurement current with
high frequency oscillations removed \footnote{This low pass filter
is far from ideal. It is possible to design low-pass filters with
much finer frequency selection properties (e.g. Butterworth
filters) \cite{sig_sys}, and we expect schemes using such filters
to perform better than this simpler version.}. The filter
parameters $r$ and $T$ determine the decay rate and length of the
filter, respectively, and $\mathcal{N} = \frac{2\kappa}{r}(1-e^{-rT})$ serves to
normalize $R(t)$ such that it is centred around $\pm 1$.

We will use the signature of this smoothed measurement signal to
infer the state of the system and thus to condition the feedback.
Explicitly, the form of the feedback conditioning current is
\begin{equation}
G(t) =\left\{ \begin{array}{ll}
    R(t) & \textrm{if $R(t)<0$} \\
  0 & \textrm{otherwise}
\end{array} \right.
\end{equation}
Thus, we describe the behaviour of the system \textit{with}
feedback using
\begin{eqnarray} \label{eq:oq_drho_fb} d\rho_c(t) &=&
\gamma\mathcal{D}[X]\rho_c(t)dt +
\kappa\mathcal{D}[Z]\rho_c(t)dt \nn \\
&&+ \sqrt{\kappa}\mathcal{H}[Z]\rho_c(t)dW(t) \nn \\ &&- i\lambda
G(t)[X,\rho_c(t)]dt
\end{eqnarray}
where $\lambda$ is the maximum feedback strength. 

Clearly this feedback conditioning current is non-Markovian (and
non-linear). As mentioned above, this makes the Markovian
simplification impossible, and therefore the most direct route to
evaluating this error correction protocol is numerical simulation.
This is done in section \ref{sec:sims}.

We note at this point that there are several open parameters in
the SME (\ref{eq:oq_drho_fb}). These parameters are the following:
\begin{enumerate}
  \item $\gamma$ - This is the error rate and is largely out of the experimenter's
  control.
  \item $r$ - The decay rate of the smoothing filter. Large values of $r$ yield responsive measurement currents, while small values of $r$ introduce more delay but make the processed measurement current smoother. We expect there to be some optimal value of $r$ that achieves a trade-off between responsiveness and smoothing ability.
 
 $r$ is intimately connected to the other filter parameter appearing in 
(\ref{eq:oq_filt}): $T$, the size of the filter's memory, which determines how many measurements from the past the filter uses in its calculations. We will choose $T$ to be large enough so that the decaying exponential filter is not truncated prematurely. A $T$ that is some large enough multiple of the filter's time constant, $1/r$, would be ideal. Since these parameters are dependent on each other, we will only consider one of them ($r$) to be free.
  
  \item $\lambda$ - The maximum strength of the feedback
  Hamiltonian. The value of this parameter is determined by the
  physical apparatus and the method of feedback. We expect the
  performance of the protocol to improve with $\lambda$, because
  increasing $\lambda$ increases the range of feedback strengths
  available.

  \item $\kappa$ - A parametrization of the measurement strength used
  in measuring the stabilizer generators. The larger $\kappa$ is
  the more information we gain from these measurements, and thus we
  expect the performance to improve with increasing $\kappa$.

\end{enumerate}
In summary, we have three parameters to control - one filter
parameter, one feedback parameter, and one measurement
parameter.  We expect there to be a region in this parameter
space where this error control scheme will perform optimally. We will investigate this issue using simulations.

\subsection{Example: Bit flip correction}
\label{sec:scheme_bitflips} This example is similar to the toy model
above but looks at a more realistic error control situation. We will
describe the dynamics of a continuous error correction scheme designed
to protect against bit flips using the three qubit bit flip code of
section \ref{sec:ex_bitflipcode}.

The measurement currents and SME of the system before the application
of feedback are
\begin{eqnarray}
\label{eq:tq_drho} d\rho_c(t) &=&
\gamma(\mathcal{D}[XII]+\mathcal{D}[IXI]+\mathcal{D}[IIX])\rho_c(t)
dt \nn \\ &&+ \kappa(\mathcal{D}[ZZI]+\mathcal{D}[IZZ])\rho_c(t)
dt \nn
\\ &&+ \sqrt{\kappa}(\mathcal{H}[ZZI]dW_1(t) \nn \\
&& ~~~~ + \mathcal{H}[IZZ]dW_2(t))\rho_c(t) \\
\label{eq:dq1} dQ_1(t) &=& 2\kappa\langle ZZI \rangle_c(t) dt +
\sqrt{\kappa}
dW_1(t) \\
\label{eq:dq2} dQ_2(t) &=& 2\kappa\langle IZZ \rangle_c(t) dt +
\sqrt{\kappa} dW_2(t),
\end{eqnarray}
where $\gamma$ is the error rate for each qubit, and $\kappa$ is
the measurement strength. We will assume that the errors on
different qubits are independent and occur at the same error rate,
and also that the measurement strength is the same for both
stabilizer generators. (The assumption of identical rates is made for
simplicity and can be removed.)

Now, as detailed in table \ref{tab:tq_table}, the measurements of
$ZZI$ and $IZZ$ reveal everything about the errors. However, as in
the toy model, we must smooth the measurement currents in order to
gain reliable error information. Therefore, the steps involved in
the error correction scheme are the following:
\begin{enumerate}
\item Smooth the measurement currents using the following filter:
\begin{equation}
\label{eq:tq_filter} R_i(t) = \frac{1}{\mathcal{N}}\int_{t-T}^t
e^{-r(t-t')}dQ_i(t') ~~~~~i=1,2
\end{equation}
The definition of this filter is analogous to (\ref{eq:oq_filt}).
\item Depending on the signatures of $R_1(t)$ and $R_2(t)$ apply
the appropriate feedback Hamiltonian. That is,
\begin{enumerate}
\item If $R_1(t)<0$ and $R_2(t)>0$, apply $XII$. \item If
$R_1(t)>0$ and $R_2(t)<0$, apply $IIX$. \item If $R_1(t)<0$ and
$R_2(t)<0$, apply $IXI$. \item If $R_1(t)>0$ and $R_2(t)>0$, do
not apply any feedback.
\end{enumerate}
These conditions translate into the following feedback
conditioning currents:
\begin{eqnarray}
G_1(t) =\left\{ \begin{array}{ll}
    R_1(t) & \textrm{if $R_1(t)<0$ and $R_2(t)>0$} \\
  0 & \textrm{otherwise}
\end{array} \right. \\
G_3(t) =\left\{ \begin{array}{ll}
    R_2(t) & \textrm{if $R_1(t)>0$ and $R_2(t)<0$} \\
  0 & \textrm{otherwise}
\end{array} \right. \\
G_2(t) =\left\{ \begin{array}{ll}
    R_1(t) & \textrm{if $R_1(t)<0$ and $R_2(t)<0$} \\
  0 & \textrm{otherwise}
\end{array} \right.
\end{eqnarray}
\end{enumerate}
Under this scheme, the SME describing the system dynamics
\textit{with} feedback becomes
\begin{eqnarray}
\label{eq:tq_drho_fb} d\rho_c(t) &=&
\gamma(\mathcal{D}[XII]+\mathcal{D}[IXI]+\mathcal{D}[IIX])\rho_c(t)
dt \nn \\ &&+ \kappa(\mathcal{D}[ZZI]+\mathcal{D}[IZZ])\rho_c(t)
dt \nn
\\ &&+ \sqrt{\kappa}(\mathcal{H}[ZZI]dW_1(t) + \mathcal{H}[IZZ]dW_2(t))\rho_c(t) \nn \\
&&- i\lambda(G_1(t)[XII, \rho_c(t)] + G_2(t)[IXI, \rho_c(t)] \nn
\\&&~~~+ G_3(t)[IIX, \rho_c(t)])dt
\end{eqnarray}
where $\lambda$ is the maximum feedback strength, which is assumed for simplicity to be the same for all the feedback Hamiltonians.

Again, the non-Markovian feedback signals make numerical
simulation the most direct method of solution of this SME.

Also, it is worth noting that even though we gain error information from the signatures of the stabilizer generators, we do not have to wait until the smoothed measurement signals, for example (\ref{eq:tq_filter}), fall below zero before turning on feedback. The feedback conditioning signals, $G_i(t)$, can be made non zero as soon as we recognize that the smoothed measurement signals are changing sign. That is, the feedback mechanism can be turned on as soon as we see a significant shift in the stabilizer generator measurements from their error free value: one. We can state this `significant' shift more precisely as a change of more than $n$ standard deviations from the mean value of one, for $n$ sufficiently large. Thus the choice of $n$ depends on the signal-to-noise ratio of the smoothed measurement currents, and therefore on the parameters $r$ and $\kappa$.

\section{Simulation Results}
\label{sec:sims} As a way of evaluating the performance of the
general error control scheme using weak measurements and feedback,
we numerically solved the SMEs described in the two examples of
section \ref{sec:scheme}. A comparison of the SMEs
(\ref{eq:oq_drho_fb}) and (\ref{eq:tq_drho_fb}) shows that the one
qubit toy model has all the free parameters of the full three
qubit code, and therefore is a good model on which to explore the
parameter space formed by $r, \lambda,$ and $\kappa$. This is
useful because the smaller state space of the toy model makes
simulating it far more computationally tractable than simulating the bit-flip
correction example.

\subsection{The toy model}
\label{sec:sims_toy} We chose to simulate the dynamics of
(\ref{eq:oq_drho_fb}) by way of an associated stochastic
Schr\"{o}dinger equation (SSE) for two reasons: (i) it is less
computationally intensive, (ii) it allows us to look at individual
trajectories of the system if desired. The form of this associated
SSE is as follows:
\begin{widetext} \begin{equation} \label{eq:oq_sse_fb} d\ket{\psi_c(t)} = dN(t)(X\ket{\psi_c(t)} - \ket{\psi_c(t)}) ~+~
\sqrt{\kappa}~dW(t)(Z-\langle Z \rangle(t))~\ket{\psi_c(t)} ~-~
\frac{\kappa}{2}(1 - \langle Z \rangle(t)Z)^2 ~\ket{\psi_c(t)}dt
~-~ i\lambda G(t)X ~\ket{\psi_c(t)}dt
\end{equation}
\end{widetext}
where $dN(t)$ is a point process increment (in the number of
errors) described by
\begin{eqnarray}
\label{eq:dN} dN(t)^2 &=& dN(t) \\
E[dN(t)] &=& \gamma dt.
\end{eqnarray}
That is, $dN$ is a random variable that is either 0 or 1 at each
time step, and is distributed according to the error rate
$\gamma$. A graph of the process $dN(t)$ would be a sequence
of Poisson distributed (with parameter $\gamma$) spikes.
In the language of quantum trajectories, this SSE is simply one
possible unravelling of the SME (\ref{eq:oq_drho_fb}).

The SSE was solved using Euler numerical integration with time
steps $dt=10^{-4}$. When ensemble averages were required --- that
is, when we were interested in the behaviour of $\rho_c(t)$ --- 600
trajectories were averaged over. To evaluate the performance of
the protocol, we used the \textit{codeword fidelity}: $F(t) =
\bra{\psi(0)}\rho(t)\ket{\psi(0)}$. Here, $\ket{\psi(0)}$ is the
initial state of the system, which is taken to be $\ket{0}$
unless otherwise specified.

\begin{figure}[h!]
\includegraphics[scale=0.45]{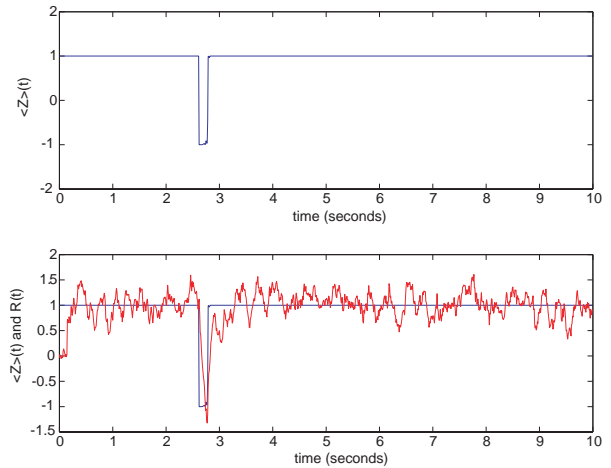}
\caption{A sample trajectory of the one qubit "code" with
feedback. The top graph just shows the expectation value of $Z$,
and the bottom graph shows expectation value of $Z$ and the
filtered signal $R(t)$.} \label{fig:oq_traj}
\end{figure}

Figure \ref{fig:oq_traj} shows a sample trajectory from the one
qubit simulation. The figure shows the expectation value of the
$Z$ measurement as a function of time and also
the superimposed filtered measurement signal, $R(t)$. The
transitions (of expectation value of $Z$) to $-1$ are due to
errors, and the transitions back to $+1$ are due to feedback
correction.

We used this toy model primarily to gain insight into the choice
of parameters that lead to optimal error correction. The
conclusions drawn from exploring the parameter space using this
one qubit simulation are the following:
\begin{enumerate}\item The decay rate of
the filter, $r$, should be determined by the strength of the feedback,
$\lambda$. That is, given a strong feedback Hamiltonian, it is
necessary to have a responsive conditioning current; one with
little memory. \item As expected, the larger the measurement
strength $\kappa$, the better the protocol performs. \item
Performance also improves as $\lambda$, the feedback strength, is
increased. This is to be expected because as $\lambda$ is
increased, the range of the strength of the feedback Hamiltonian
increases leading to a greater degree of control.
\item The interplay between the two processes -- measurement and feedback -- must be considered. In particular, larger values of $\kappa$ will yield better performance only if these values are not too much larger than the value of $\lambda$. That is, if the measurement strength is much stronger than the feedback strength, the measurement process disrupts the feedback correction process and makes it ineffective. Therefore the magnitude of the measurement strength should be less than, or of the same order of magnitude, as the feedback Hamiltonian strength.   
\end{enumerate}

Given the strong dependence between parameters that these
one qubit simulations identify, there are really only three free
parameters in the system: $\kappa, \lambda$ and $\gamma$.
Since the last is out of the experimenter's control, there remain
two controllable parameters. In practice, neither of these
parameters, the measurement strength or the feedback strength, are
completely configurable. The physical implementation scheme
typically limits the range of these parameters, and in section
\ref{sec:real} we shall see whether the practical ranges for one
particular implementation allows for error control via this
feedback scheme.

It is instructive to note that the free parameters of the protocol
are all physical parameters. That is, the optimal operating regime
of the protocol is defined by the system's physical features
rather than the introduced filter. Therefore, it is possible to design a filter that allows the protocol to perform optimally for a given set of physical parameters ($\kappa$ and $\lambda$).

\subsection{ Three qubit code simulation }
\label{sec:sims_threeq} Now we move on to the simulation of the
three qubit bit flip code. This simulation behaves in much the
same way as the one qubit version, but with one key difference:
for the one qubit `code', a double error event -- where an error
occurs on the qubit before we have corrected the last error -- is
not too damaging: in this case, the error correcting feedback
mechanism detects a traversal back into the `codespace' and thus
stops correcting. In the three qubit code, this situation is a
little more complicated. Let us consider the situation in which a second error happens while a previous error is being corrected. 
If this second error happens to be on the same qubit
as the one being corrected, then in consonance with the one qubit
`code', it is not too damaging. However, if the second error is on
one of the two qubits not being corrected, an
irrecoverably damaging event occurs, because in this case the stabilizer
measurements cease to provide accurate information about the error
location, and the protocol's `corrections' actually introduce
errors.

This identifies a key consideration in any continuous, feedback based
error correction scheme. The finite duration of the detection and
correction window means that we must choose our parameters with
this finite window small enough that the probability of an error
we cannot correct (in this case, two errors on different qubits) is
negligible. This is analogous to choosing the detection-correction
intervals in the discrete error control case to be small enough to
avoid uncorrectable errors.

The SSE that describes the dynamics of the three qubit error
correction scheme is
\begin{widetext}
\begin{eqnarray}
\label{eq:tq_sse_fb} d\ket{\psi_c(t)} &=&
dN_1(t)(XII\ket{\psi_c(t)} - \ket{\psi_c(t)}) ~+~
dN_2(t)(IXI\ket{\psi_c(t)} - \ket{\psi_c(t)}) ~+~
dN_3(t)(IIX\ket{\psi_c(t)} - \ket{\psi_c(t)}) \nn \\
&+& \sqrt{\kappa}~dW_1(t)(ZZI-\langle ZZI
\rangle(t))~\ket{\psi_c(t)} ~+~
\sqrt{\kappa}~dW_2(t)(IZZ-\langle IZZ \rangle(t))~\ket{\psi_c(t)} \nn \\
&-& \frac{\kappa}{2}(1 - \langle ZZI \rangle(t)ZZI)^2
~\ket{\psi_c(t)}dt ~-~ \frac{\kappa}{2}(1 - \langle IZZ
\rangle(t)IZZ)^2 ~\ket{\psi_c(t)}dt \nn \\
&-& i\lambda G_1(t)XII ~\ket{\psi_c(t)}dt ~-~ i\lambda G_2(t)IXI
~\ket{\psi_c(t)}dt ~-~ i\lambda G_3(t)IIX ~\ket{\psi_c(t)}dt.
\end{eqnarray}
\end{widetext}
This SSE is of course an unravelling of SME (\ref{eq:tq_drho_fb}),
and all parameters are defined as for that equation.

As in the one qubit case, we solved this differential equation
using an Euler method with timesteps $dt=10^{-4}$. Again, ensemble
averages were done over 600 trajectories when needed. The initial
state used was $\ket{000}$, and the performance was measured using
the codeword fidelity $F_3(t) = \bra{000}\rho(t)\ket{000}$. A true
fidelity measure of the protocol performance would average over
all possible input states; however, because $\ket{000}$ is most
susceptible to bit flip errors, the fidelity we use can be
considered a worst case performance analysis.

\begin{figure}[h!]
\includegraphics[scale=0.5]{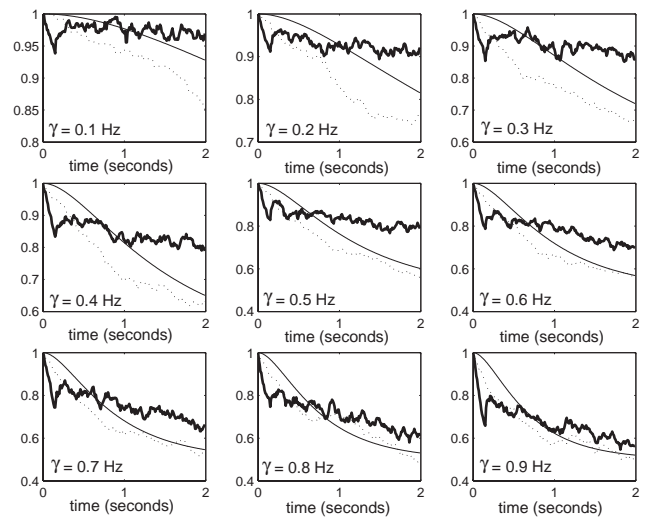}
\caption{Fidelity curves with and without error correction for
several error rates. The thick solid curve is the fidelity of the
three qubit code with error correction, $F_3(t)$ (parameters used:
$dt=10^{-4} s, \kappa=150~Hz, \lambda=150~Hz, r=20~Hz, T=1500\times dt~s$). The
dotted curve is the fidelity of one qubit without error
correction, $F_1(t)$. And the thin solid curve is the fidelity
achievable by discrete quantum error correction when the duration
between applications is \textit{t}, $F_{3d}(t)$.} \label{fig:fids}
\end{figure}

The performance of the error correction scheme using this code is
summarized by figure \ref{fig:fids}. This figure shows the fidelity
versus time curves ($F_3(t)$) for several values of error rate
($\gamma$).  Each plot also shows the fidelity curve ($F_1(t)$) for
one qubit in the absence of error correction. A comparison of these
two curves shows that the fidelity is preserved for a longer
period of time by the error correction scheme for small enough error
rates. Furthermore, for small error rates ($\gamma<0.3$) the $F_3(t)$
curve shows a vast imrpovement over the exponential decay in the absence of error
correction. However, we see that past a certain threshold error rate, the
fidelity decay even in the presence of error correction behaves
exponentially, and the two curves look very similar; past the threshold, the error correcting scheme becomes
unable to handle the errors and becomes ineffective. In fact, although not completely evident from the figure, well above the threshold the performance of the scheme becomes worse than the unprotected qubit's performance. This poor performance results from the
feedback `corrections' being so inaccurate that the feedback mechanism
effectively increases the error rate.

The third line in the plots of figure \ref{fig:fids} is of the average
fidelity achievable by discrete quantum error correction --using
the same three qubit code-- when the time between the
detection-correction operations is \textit{t}. The value of this
fidelity ($F_{3d}(t)$) as a function of time was analytically
calculated in \cite{ahn-dl},
\begin{equation}
\label{f3} F_{3d} = \frac{1}{4}(2 + 3e^{-2\gamma t} - e^{-6\gamma
t}).
\end{equation}
A comparison between $F_3(t)$ and $F_{3d}(t)$ highlights the
relative merits of the two schemes. The fact that the two curves
cross each other for large $t$ indicates that if the time between
applications of discrete error correction is sufficiently large,
then a continuous protocol will preserve fidelity better than a
corresponding discrete scheme. In fact, this comparison suggests
that a hybrid scheme, where discrete error correction is performed
relatively infrequently on a system continuously protected by a
feedback protocol, might be a viable approach to error control.

All the $F_3(t)$ curves show an exponential decay at
very early times, $t\approx 0$ to $t\approx 0.1$. This is an
artifact of the finite filter length and our specific
implementation of the protocol. In particular, our simulation does
not smooth the measurement signal until enough time has passed to
get a full buffer of measurements; that is, filtering and feedback
only start at $t=T$. Of course, this can be remedied by a more
complicated scheme that smoothes the measurement signal and
applies feedback even when it has access to fewer than $T/dt$
measurements.

\subsection{ Inefficient measurement }
We have modeled all our measurement processes as being perfect.
In reality, detectors will be inefficient and thus yield
imperfect measurement results. This ineffiency is typically represented by a parameter $\eta$ that can range from 0 to 1, where 1 denotes a perfect detector. How is this feedback protocol
affected by non-unit efficiency detection?

To examine this question, we simulated the three qubit code with
inefficient detection. The evolution SME and the measurement
currents in the presence of inefficient detection are as follows:
\begin{eqnarray}
\label{eq:tq_drho_w_eta} d\rho_c(t) &=&
\gamma(\mathcal{D}[XII]+\mathcal{D}[IXI]+\mathcal{D}[IIX])\rho_c(t)
dt \nn \\ &&+ \kappa(\mathcal{D}[ZZI]+\mathcal{D}[IZZ])\rho_c(t)
dt \nn
\\ &&+ \sqrt{\kappa\eta}(\mathcal{H}[ZZI]dW_1(t) + \mathcal{H}[IZZ]dW_2(t))\rho_c(t) \nn \\
&&- i\lambda(G_1(t)[XII, \rho_c(t)] + G_2(t)[IXI, \rho_c(t)] \nn
\\&&~~~+ G_3(t)[IIX, \rho_c(t)])dt \\
\label{eq:dq1_w_eta} dQ_1(t) &=& 2\kappa\sqrt{\eta}\langle ZZI
\rangle_c(t) dt + \sqrt{\kappa}
dW_1(t) \\
\label{eq:dq2_w_eta} dQ_2(t) &=& 2\kappa\sqrt{\eta}\langle IZZ
\rangle_c(t) dt + \sqrt{\kappa} dW_2(t)
\end{eqnarray}
where $0 < \eta \leq 1$ is the measurement efficiency, and all
other quantities are the same as in equations (\ref{eq:tq_drho})
and (\ref{eq:tq_drho_fb}).

\begin{figure}[h!]
\includegraphics[scale=0.45]{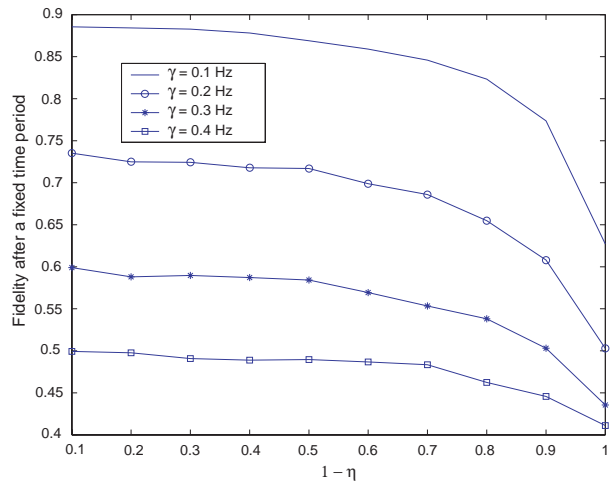}
\caption{Average fidelity after a fixed amount of time as a function of 1-efficiency for several error rates (parameters used: $dt=10^{-4} s, \kappa=50~Hz, \lambda=50~Hz, r=10~Hz, T=1500\times dt~s$). } \label{fig:eta_sims} \end{figure}

The results of these simulations are summarized by figure
\ref{fig:eta_sims}. The decay of fidelity with decreasing $\eta$
indicates that inefficient measurements have a negative effect on the
performance of the protocol as expected. However, the slope of the
decay is very small--- in particular, the graph does not exponentially decay as do Markovian feedback protocols--- and this suggests that this
protocol has a certain tolerance to inefficiencies in
measurement. This is reasonable because the filtering process in the
protocol has the effect of improving the quality of the measurements
and thus negating some of the ill effects of the inefficient
measurements. Also, as in the full state estimation protocol of
\cite{ahn-dl}, because the feedback conditioning current is a function
of a measurement record history ---as opposed to just the current
measurement--- errors induced by inefficient measurement tend not to
be so damaging. Here we see the true strength of this error correction
scheme: it combines the robustness of a state estimation based
feedback protocol with the practicality of a Markovian feedback
protocol.

\section{Links to experiment}
\label{sec:real} In this section we study the possibility of applying this error correction technique to a particular quantum computing architecture. 

\subsection{Solid-state quantum computing with RF-SET readout}
Several schemes for solid-state quantum computing have been
proposed \cite{kane, qdots, qdots-cnot, cooperpair}. These use the
charge or spin degree of freedom of single particles to represent
logical qubits, and measurement involves probing this degree of
freedom.

Here we examine the weak measurement of one such proposal that
uses coherently coupled quantum dots (CQDs) and an electron that
tunnels between the dots\cite{Hollenberg03}. The dots are formed
by two P donors in Si, separated by a distance of about 50nm.
Surface gates are used to remove one electron from the double
donor system leaving a single electron on the P-P$^+$ system.
This system can be regarded as a double well potential.  Surface
gates can then be used to control the barrier between the wells as
well as the relative depth of the two wells.  Using surface gates,
the wells can be biased so that the electron can be well localised
on either the left $|L\rangle$ or the right $|R\rangle$ of the
barrier. These (almost) orthogonal localised states are taken as
the logical basis for the qubit, $\ket{0}=\ket{L},\ \
\ket{1}=\ket{R}$. It is possible to design the double well
system so that, when the well depths are equal, there are only two
energy eigenstates below the barrier. These states are the
symmetric ground state $\ket{+}$ and the antisymmetric first
excited state $\ket{-}$. A state localised on the left (right)
of the barrier is then well approximated as a linear superposition
of these two states,
\begin{eqnarray}
|L\rangle & = & \frac{1}{\sqrt{2}}(|+\rangle+|-\rangle)\\
|R\rangle & = & \frac{1}{\sqrt{2}}(|+\rangle-|-\rangle)
\end{eqnarray}
An initial state localised in one well will then tunnel to the
other well at the frequency $\Delta=(\epsilon_+-\epsilon_-)/\hbar$
where $\epsilon_\pm$ are the two energy eigenstates below the
barrier.

The Pauli matrix, $Z=|L\rangle\langle L|-|R\rangle\langle R|$, is
diagonal in this localised state basis.  The Hamiltonian for the system
can be well approximated  by
\begin{equation}
H=\hbar\frac{\omega(t)}{2}Z+\hbar\frac{\Delta(t)}{2} X
\label{eq:hamiltonian}
\end{equation}
where $X=|L\rangle\langle R|+|R\rangle\langle L|$. Surface gates
control the relative well depth  $\hbar\omega(t)$ (a bias gate
control) and the tunnelling rate $\Delta(t)$, (a barrier gate
control) which are therefore  time dependent. For non-zero bias the
energy gap between the ground state and the first excited state is
$E(t)=\hbar\sqrt{\omega(t)^2+\Delta(t)^2}$. Further details on the
validity of this Hamiltonian and how well it can be realised in
the PP$^+$ in Si system can be found in \cite{Barrett03}.

A number of authors have discussed the sources of decoherence in a charge qubit system such as
this one\cite{Barrett03,Fedichkin03,Hollenberg03}. For appropriate
donor separation, phonons can be neglected as a source of
decoherence. The dominant sources of decoherence then arise from
fluctuations in voltages on the surface gates controlling the
Hamiltonian and electrons moving in and out of trap states in the
vicinity of the dot. This latter source of decoherence is expected
to occur on a longer time scale and is largely responsible for
$1/f$ noise in these systems. In any case both sources of
decoherence can be modelled using the well known spin-boson model \cite{weiss}.
The key element of this model for the discussion here is that the
interaction energy between the qubit and the reservoir is a
function of $Z$.

If the tunnelling term proportional to $\Delta(t)X$ in
Eq. (\ref{eq:hamiltonian}) were not present, decoherence of this kind
would lead to pure dephasing. However, in a general single qubit
gate operation, both dephasing and bit flip errors can arise in
the spin-boson model. We can thus use the decoherence rate
calculated for this model as the bit flip error rate in our
feedback error correction model. We will use  the result from the
detailed model of Hollenberg et al. \cite{Hollenberg03} for a
device operating at 10K, and set the error rate $\gamma=1.4 \times
10^6\mbox{s}^{-1}$. This rate could be made a factor of ten smaller by operating at lower temperatures and improving the electronics controlling the gates. 

We now turn to estimating the measurement rate , $\kappa$, for the
PP$^+$ system. In order to readout the qubit in the logical basis
we need to distinguish a single electron in the left or the right
well quickly and with high probability of success (efficiency).
The technique of choice is currently based on radio frequency
single electron transistors (RF-SET)\cite{Schoelkopf98}. We will
use the twin SET implementation of Buehler et al.
\cite{BuehlerB03}.

In an RF SET the Ohmic load in a tuned tank circuit comprises a single
electron transistor with the qubit acting as a gate bias.  The two
different charge states of the qubit provide two different bias
conditions for the SET, producing two different resistive loads, and
thus two levels of power transmitted through the tank circuit. The
electronic signal carries a number of noise components: for example,
the Johnson-Nyquist noise of the circuit, random changes in the SET
bias conditions due to fluctuating trap states in the SET, etc.  The
measurement must be operated in such a way that the charge state of
the qubit can be quickly discerned as a departure of the signal from
some fiducial setting, despite the noise. Clearly it takes some
minimum time interval, $t_M$, to discriminate a qubit signal change
from a random noisy fluctuation. We need to keep the measurement time
as short as possible. However if the measurement time is too short,
one may mistake a large fluctuation, due to a non-qubit based change
in bias conditions, for the real signal. In other words one may
mistake a 1 for a 0, and vice versa. The probability of this happening
is the efficiency of the measurement, $\eta(t_M)$, which depends on
the measurement time.  The key performance parameters are (i) the
measurement time, $t_M$, and (ii) the efficiency $\eta(t_M)$. An
additional parameter that is often quoted is the minimum charge
sensitivity per root hertz, $S$. Given $t_M$, $S$ determines a minimum
change in the charge, $\Delta q$, that can be seen by the RF-SET at a
given bias condition.  In \cite{BuehlerB03}, a measurement time of
$t_M=6\times 10^{-6}\mbox{s}$ was found for a signal of $\Delta
q=0.2e$ and an efficiency of $10^{-6}$.  We now need to relate this
measurement time to the measurement decoherence rate parameter,
$\kappa$, of our ideal feedback model.

If the measurement were truly quantum limited (that is to say, the
signal-to-noise ratio is determined only by the decoherence rate
$\kappa$), the inverse measurement time would be of the same order of
magnitude as the decoherence rate (see \cite{Goan03}). The measurement
described in Buehler et al.\cite{BuehlerB03} will almost certainly not
be quantum limited. However, here we will assume the measurement to be
quantum limited, so as to obtain a lower limit to the measurement
decoherence rate. Thus we take $\kappa=10^6\mbox{s}^{-1}$.

We next need to estimate typical values for the feedback strength.
From Eq.  (\ref{eq:oq_drho_fb}) we see that the feedback Hamiltonian is
proportional to an $X$ operator. In the charge qubit example, this
corresponds to changing the tunneling rate for each of the double dot
systems that comprise each qubit.  The biggest tunneling rate
($\Delta$) occurs when the bias of the double wells makes it
symmetric. In \cite{Barrett03}, the maximum tunneling rate is about
$10^{9}$ s$^{-1}$,  for a donor separation of 40nm.  A large
tunneling rate makes for a fast gate, and thus a fast correction
operation.  Thus the maximum value of $\lambda$ can be taken to be $10^9$ s$^{-1}$.

To summarise, in the PP$^+$ based charge qubit, with RF-SET
readout, we have $\gamma\approx\kappa\approx 10^6\mbox{s}^{-1}$, and $\lambda \approx 10^{9} \mbox{s}^{-1}$.

The fact that the measurement strength and the error rate are of the
same order of magnitude for this architecture is a problem for our
error correction scheme. This means that the rate at which we gain
information is about the same as the rate at which errors happen, and
it is difficult to operate a feedback correction protocol in such a
regime. Although it is unlikely that the measurement rate could be
made significantly larger in the near future, as mentioned above it is
possible that the error rate could be made smaller by improvements in
the controlling electronics. Thus it is interesting to consider how
low the error rate would have to be pushed before our error control
scheme becomes effective. To answer this question we ran the three
qubit bit flip code simulation using the parameters stated above and
lowered the error rate until the error control performance was
acceptable. We found that the fidelity after 1ms could be kept above
0.8 on average if the error rate, $\gamma$, is below $10^2$ s$^{-1}$
(with $\kappa=10^6$ s$^{-1}$, and $\lambda=10^7$ s$^{-1}$). So we see
that a difference in order of magnitude of four between the
measurement and feedback strengths, and the error rate, is about what
this protocol (using the three qubit code) requires for reasonable
performance. That is, we require
\begin{equation}
\frac{\kappa}{\gamma} \approx \frac{\lambda}{\gamma} \approx 10^4
\end{equation}
Of course, depending on the performance requirements this ratio may be larger or smaller. Also, a full optimization of the filter used in the scheme is likely to drive this ratio down by up to an order of magnitude.

We can compare the requirements of the three-qubit code with the one-qubit version. Given the same measurement and feedback parameters ($\kappa=10^6$ s$^{-1}$, $\lambda=10^7$ s$^{-1}$), the one-qubit `code' can keep the fidelity above 0.8 after 1ms when $\kappa/\gamma \approx \lambda/\gamma \approx 10$. That is, only one order of magnitude difference is required between the error rate and the measurement and feedback rates. This suggests that a key issue with feedback based error correction schemes is \emph{scalability}. The ratio between measurement and feedback rates and error rate has to increase along with the error correcting code size (in qubits). 

\section{Discussion and Conclusion}
\label{sec:conc} We have described a practical scheme for
implementing error correction using continuous measurement and
Hamiltonian feedback. We have demonstrated the validity of the
scheme by simulating it for a simple error correction scenario.

As the simulations show, this error control scheme can be made
very effective if the operational parameters (measurement
strength, feedback strength, filter parameters) are well matched
to the error rate of a given sytem. At the same time, the
scheme uses relatively modest resources and thus is easy to
implement, as well as being robust in the face of measurement
inefficiencies.

From a quantum control perspective, an interesting feature of this
protocol is the encoding. That is, despite using state estimate
feedback, the protocol requires little side processing due to the fact
that instead of a full state estimate, it uses a coarse-grained state
estimate naturally suggested by the encoding. In control theory terms, this
simplification is a result of the specific choice of control state
space (what to observe and control); a choice dictated by the
stabilizer encoding and measurements. It would be interesting to
examine the general conditions under which an encoding is available
that allows for practical, \textit{efficient} state estimate feedback
control.

The possibility of using continuous error correction in
combination with its discrete counterpart is an interesting
possibility. Such a scheme has the potential to significantly
improve the stability of quantum memories, and the implications of
such a combination scheme for fault tolerance would be worth
investigating.

We also studied a solid-state quantum computing architecture with RF-SET readout and the feasibility of implementing this error correction protocol on it. Although the measurement and feedback rates currently possible on this architecture do not allow for error correction via this feedback scheme with the intrinsic error rate, it is foreseeable that as the controlling technology improves, this error control scheme will become possible on this architecture. From numerical simulations, we found the approximate parameter regime where the three qubit code using this scheme becomes effective -- that is, exactly how much improvement is necessary before the scheme becomes feasible. It would be interesting to investigate this further and explore more rigorously how values of $\kappa/\gamma$ and $\lambda/\gamma$ dictate protocol performance as well as the exact dependency of these parameter ratios on the code size. Such an investigation will be crucial in addressing the issue of scalability of this error control scheme.

\section{Acknowledgements}
\label{sec:acks} We gratefully acknowledge the support of the
Australian Research Council Centre of Excellence in Quantum
Computer Technology. Part of this research was done using the
resources of the Visualisation and Advanced Computing Laboratory
at the University of Queensland. MS would like to thank Mr. James
Lever for his assistance and support at this facility. CA is grateful
for the hospitality of the Centre for Quantum Computer Technology at
the University of Queensland and acknowledges support from an
Institute for Quantum Information fellowship.

\bibliography{ecwfv6}

\end{document}